\documentclass[conference]{IEEEtran}
\IEEEoverridecommandlockouts

\def\BibTeX{{\rm B\kern-.05em{\sc i\kern-.025em b}\kern-.08em
    T\kern-.1667em\lower.7ex\hbox{E}\kern-.125emX}}

\usepackage{soul} 
\usepackage{macros}
\usepackage{xspace}
\usepackage{graphicx}
\usepackage{kotex} 
\usepackage{comment} 
\usepackage{subcaption}
\usepackage{amsmath, amsfonts}
\usepackage{algorithm}
\usepackage{textcomp}
\usepackage{balance} 
\usepackage{algpseudocode} 
\usepackage{mathtools}
\usepackage{makecell}
\usepackage{tikz}
\usepackage{xcolor} 
\usepackage{booktabs} 
\usepackage{amssymb} 
\usepackage{cite} 
\usepackage{fancyhdr}
\usepackage{hyperref} 
\usepackage{tabularx}

\usepackage{booktabs} 
\usepackage{multirow} 

\begin{document}

\title{Throughput-Oriented LLM Inference via KV-Activation Hybrid Caching with a Single GPU}

    



\author{
    \IEEEauthorblockN{
        Sanghyeon Lee,
        Hongbeen Kim,
        Soojin Hwang,
        Guseul Heo,
        Minwoo Noh,
        Jaehyuk Huh
    }
    \IEEEauthorblockA{
        School of Computing, KAIST, Daejeon, Republic of Korea \\
        \{leesh6796, hbkim, sjhwang, gsheo, mwnoh\}@casys.kaist.ac.kr, jhhuh@kaist.ac.kr
    }
}

\maketitle

\begin{abstract}

Large language models (LLMs) with growing model sizes use many GPUs to meet memory capacity requirements, incurring substantial costs for token generation.
To provide cost-effective LLM inference with relaxed latency constraints, recent studies proposed to expand GPU memory by leveraging the host memory.
However, in such host memory offloading, the host-GPU interconnect becomes a performance bottleneck for the transfer of weights and key-value (KV) cache entries, causing the underutilization of GPUs.
To address the challenge, we introduce \capture, an LLM inference system with KV-activation hybrid caching. The activation cache stores activation checkpoints instead of keys and values during intermediate inference stages, requiring half of the memory capacity compared to the conventional KV cache. 
While model parameters are transferred to GPU from host memory, idle GPU resources can be utilized for the partial recomputation. 
To balance the latency of activation recomputation and parameter loading, our KV-activation hybrid caching scheme determines the optimal ratio between KV and activation caches to manage both recomputation time and data transfer times.
\capture achieves 2.19$\times$ throughput improvement over the state-of-the-art prior work for offloading both model weights and KV cache.


\end{abstract}


\section{Introduction}


Large language models (LLMs) based on the Transformer decoder architecture require substantial capacity of GPU memory to store both model weights and a Key-Value cache (KV cache), which holds the keys and values of the prompt and previously generated tokens for every layer~\cite{opt, llama2}. 
The KV cache accounts for a significant portion of the GPU memory, and its size grows as the token sequence length or batch size increases. The substantial memory capacity requirements of both the KV cache and model weights have increased the cost of LLM services, as multiple GPUs are required for real-time inference~\cite{orca, pagedattention}.
However, in many non-interactive scenarios such as dataset evaluations or large-scale text analysis~\cite{summary-plz}, higher latency is acceptable if it leads to the better cost efficiency. 
Leveraging host CPU memory—which offers large capacity with additional latency overhead—is a promising solution for these scenarios.
Such \textit{host memory offloading} can significantly reduce the cost of LLM processing at the expense of increased latency~\cite{deepspeed-inference}. 
FlexGen showed that both model weights and KV cache can be offloaded to host memory, enabling a single-GPU LLM inference~\cite{flexgen}. 


The cause of main performance bottleneck of LLM inference with \textit{host memory offloading} is the limited communication bandwidth between host memory and the GPU. 
Since both the model weights and KV cache must be transferred to the limited-capacity GPU memory, the GPU often becomes idle waiting for the next operands to arrive from host memory—even when communication and computation are overlapped.


\begin{figure}[t]
    \vspace{-4ex}
    \centering
    \includegraphics[width=0.75\linewidth]{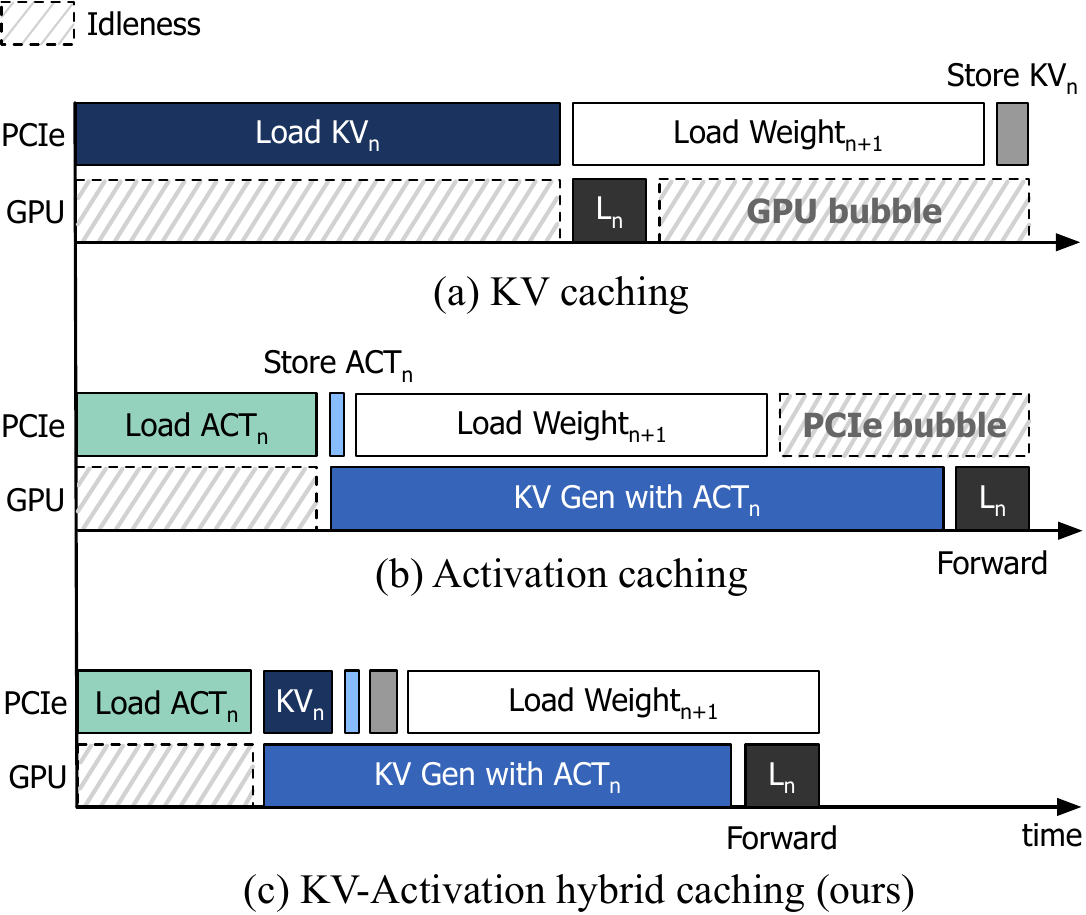}
    \caption{Timeline comparison: (a) KV caching causes GPU idle time. (b) Activation caching causes PCIe idle time. (c) Our hybrid caching eliminates both GPU and PCIe idle time by balancing transfers and computations.}
    \vspace{-2ex}
    \label{fig:caching-strategy}
\end{figure}


To address the memory and communication challenges of host memory offloading, this study proposes a novel {\it KV-activation hybrid cache} scheme based on activation caching. 
In activation caching, unlike token recomputation that relies on a full prefill step, input activation values of all layers are stored in memory. Computing the KV cache of a layer from the activation values requires a modest computation capacity, compared to complete recomputation from token IDs. Keeping activation checkpoints instead of keys and values reduces the memory footprint and communication traffic of cached contexts by half.
The activation caching maintains the same result as the conventional KV caching.

Although replacing the entire KV cache with the activation cache can reduce the memory usage by half, the amount of computation with activation recomputation also increases as the batch size and/or sequence length increases. Therefore, a balanced approach is necessary to determine the best ratio of KV and activation cache entries stored in the host memory. 
Figure~\ref{fig:caching-strategy} compares three approaches: (a) KV caching, (b) activation caching, and (c) \emph{KV-activation hybrid caching}. While the first choice over-consumes host-GPU interconnect bandwidth and the second one requires frequent KV recomputations, we deploy hybrid design (c) which reduces communication latency and computational overhead to maximize throughput by balancing both cache types.
Our {\it KV-activation hybrid caching system}, called \capture, stores parts of the prior KV cache entries as activation representation, and the remaining parts as key and value representation. 
We propose an algorithm to find the best ratio of KV and activation entries to maximize GPU computation and host-GPU interconnect utilization, and a scheduling technique to pack requests into mini-batches with the best ratio of KV and activation entries. 

We implement \capture by extending vLLM with the host memory offloading capabilities for both weight and KV cache~\cite{pagedattention}. 
In evaluation with four variants of OPT models, \capture shows the throughput improvement of 2.19$\times$ over FlexGen~\cite{flexgen}, and 1.35$\times$ compared to the activation cache-only system in geometric mean.
We open-source the implementation of \capture at \url{https://github.com/casys-kaist/Capture}.





\section{Background and Related Work}
\label{sec:background}

\subsection{LLM Inference Serving}
Recent research on LLM inference serving has primarily focused on latency-sensitive, multi-turn conversational applications~\cite{cachedattention, llumnix}, where multi-GPU strategies enable real-time interaction. Approaches in this domain are designed to sustain high throughput under strict latency constraints~\cite{splitwise, distserve, sarathi, orca, pagedattention}, despite the difference of main ideas.
However, as model sizes grow larger and request concurrency intensifies, maintaining a low-latency focus under all conditions becomes prohibitively expensive in terms of infrastructure and operational costs. In contrast, non-interactive scenarios like dataset evaluation and document summarization~\cite{summary-plz} can accommodate longer processing times, making throughput and cost efficiency more critical than immediate responsiveness.

\subsection{Batched LLM Inference}
\label{subsec:batched_llm_inference}
Batched LLM inference offers a cost-efficient way for real-time processing by amortizing computational overhead across multiple requests. By grouping large numbers of queries into a single batch, it reduces redundant operations such as repeated weight loading and improves GPU utilization. For example, OpenAI’s Batch API~\cite{openai-batch} processes up to 50,000 requests per batch, delivering results within 24 hours while reducing costs by 50\% compared to real-time inference. This approach is particularly well-suited for non-interactive workloads where immediate response times are not required, such as dataset evaluations, content classification, and embedding generation.

\subsection{Host Memory Offloading}
Host memory offloading alleviates the shortage of GPU memory capacity by utilizing host CPU memory to store weights and KV cache, enabling cost-effective batched inference. For efficiency, weights and KV cache of layer $i+1$ are prefetched during the processing of layer $i$. DeepSpeed-Inference~\cite{deepspeed-inference} progressively transfers weights on demand, reducing peak GPU usage, and FlexGen~\cite{flexgen} subdivides large batches into mini-batches per layer to further mitigate memory constraints. Recent methods also employ approximation techniques to reduce data transfers~\cite{infinigen, powerinfer}, but these can introduce errors that are detrimental to tasks that require high precision.

\section{Motivation}

\begin{figure}[t]
    \centering
    \includegraphics[width=\linewidth]{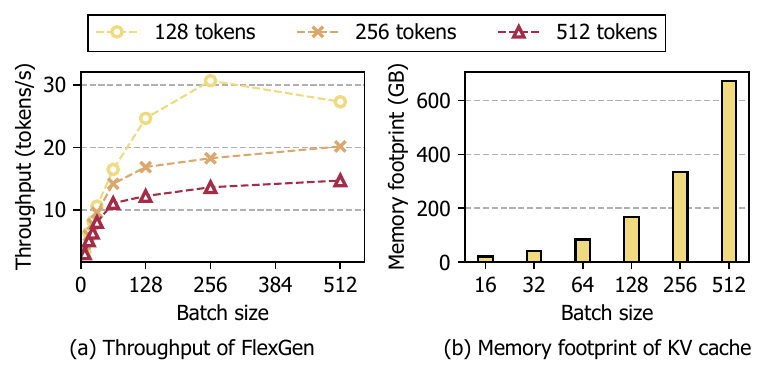}
    \caption{Performance of FlexGen~\cite{flexgen} with OPT-30B: (a) token generation throughput, and (b) KV cache memory footprint for 1024 input tokens, across varying batch sizes.}
    \label{fig:batch-centric}
\end{figure}

\begin{table}[t]
    \centering
    \small 
    \begin{tabular}{c|cccccc}
        \toprule
        \textbf{Prompt} & \textbf{B=1} & \textbf{B=8} & \textbf{B=16} & \textbf{B=64} & \textbf{B=256} & \textbf{B=1,024} \\
        \midrule
        128 toks & 3.96 & 5.87 & 6.83 & 6.86 & 6.76 & 5.77 \\
        256 toks & 3.93 & 5.76 & 6.42 & 7.02 & 7.14 & 7.15 \\
        512 toks & 3.49 & 5.57 & 6.35 & 7.27 & 6.99 & 6.28 \\
        \toprule
    \end{tabular}
    \caption{Token generation throughput of PowerInfer~\cite{powerinfer} in LLaMA-70B~\cite{llama2}. B denotes batch size.}
    \label{tab:powerinfer-throughput}
\end{table}



\subsection{Challenges for Increasing the Batch Size}

\niparagraph{Limited benefits of increasing batch size.}
Increasing the batch size is a common strategy to improve generation throughput. Larger batches allow more tokens to be generated from a single transfer of weights from host memory, maximizing weight reuse and reducing the overhead of frequent data transfers. However, larger batch size does not always result in proportional gains. As shown in Figure~\ref{fig:batch-centric} (a), throughput scales linearly for batch sizes below 128 but saturates beyond this point, with minimal performance improvements as batch size increases further.

\niparagraph{Bottleneck from KV cache transfer.}
Increasing the batch size improves weight reuse, but introduces a bottleneck in attention operations because the KV cache cannot be shared across multiple requests (i.e. unable to be batched). As a result, the PCIe transfer volume scales with the sum of context lengths, limiting GPU utilization. Figure~\ref{fig:batch-centric} (a) demonstrates that throughput saturates for batch sizes above 128 in FlexGen~\cite{flexgen}. As illustrated in Figure~\ref{fig:batch-centric} (b), this saturation occurs because of the increasing amount of KV cache transfer: The size of KV cache grows linearly with batch size, reaching up to 168GB per single iteration of token generation. Such large transfers impose significant pressure on PCIe bandwidth, reducing GPU utilization to as low as 7.4\%. Even PowerInfer~\cite{powerinfer}, which loads frequently accessed weights onto GPU while keeping sparse weights on CPU, experiences throughput saturation at large batch sizes (Table~\ref{tab:powerinfer-throughput}), underscoring that escalating KV cache traffic remains the main bottleneck.



\subsection{Limitation of KV Cache Recomputation}
\label{subsec:motiv-recompute-limitation}

\begin{figure}[t]
    \centering
    \includegraphics[width=0.9\linewidth]{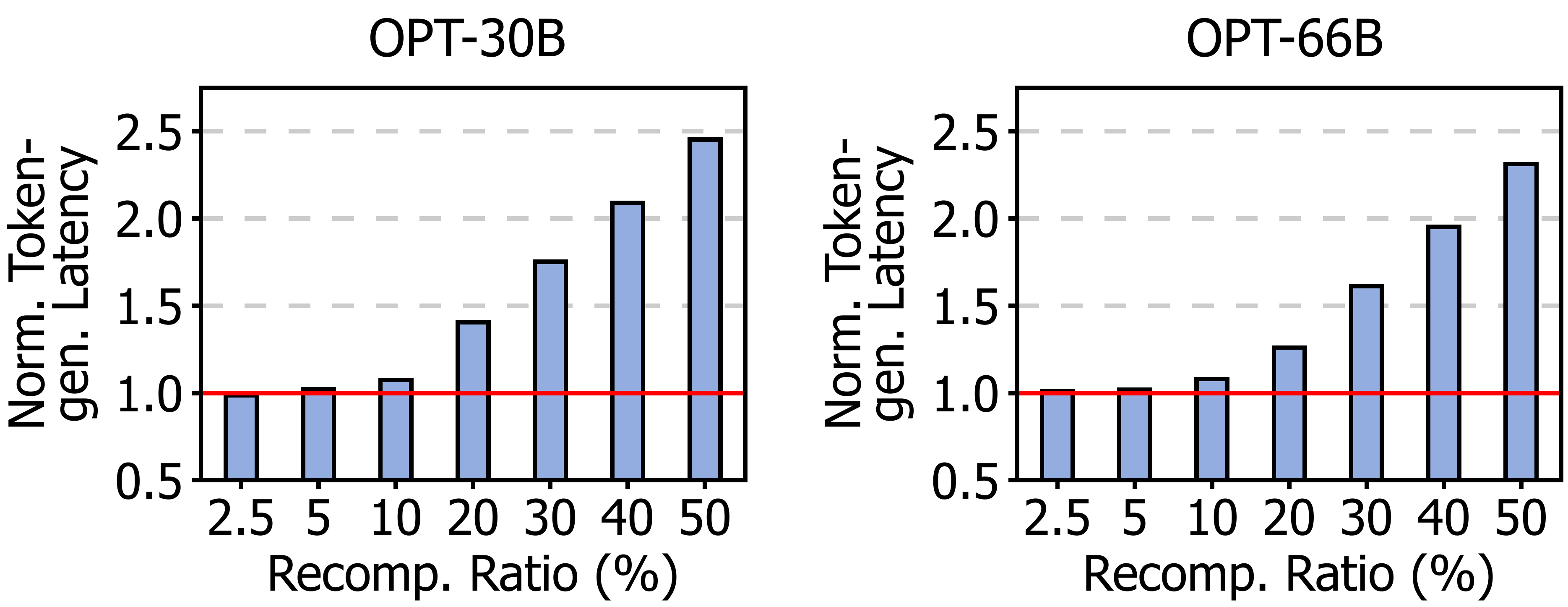}
    \caption{Token generation latency normalized to the baseline (no recomputation) for varying recomputation ratios on OPT-30B (left) and 66B (right). The red line denotes the normalization reference.}
    \vspace{-0.2in}
    \label{fig:recomp-latency}
\end{figure}

\niparagraph{Token recomputation.}
Rather than storing or transferring all KV entries, token recomputation reconstructs the cache by re-executing the prefill phase using the tokenized representations of the accumulated context (prompt + generated tokens). By eliminating the need for full KV storage or transfer, this method significantly reduces memory consumption and data movement. In principle, this enables larger batch sizes and shifts communication overhead into computation, potentially improving scalability over fully cached KV data.

\niparagraph{Limited performance gains of recomputation.}
Despite reduction of host-to-GPU data transfers, recomputation does not always yield better throughput. As illustrated in Figure~\ref{fig:recomp-latency}, with a batch size of 64 and context lengths of 1024 for the 30B model or 512 for the 66B model, a recomputation ratio of 20\% reduces data transfer overhead by 20\% but increases latency by 1.45× and 1.31×, respectively. Hence, the overhead from repeated prefill steps outweighs the savings from fewer transfers.

\niparagraph{FFNs dominate recomputation latency.}
Reconstructing KV tensors from token IDs requires not only regenerating KV entries for the target layer but also recomputing all preceding layers to reconstruct their intermediate activations. As shown in Figure~\ref{fig:cache-difference}(a), this sequential recomputation significantly inflates latency by processing each layer in order. Additionally, Figure~\ref{fig:remat-breakdown} highlights that feed-forward networks (FFNs), which generate inputs for subsequent layers, dominate execution time (see the \texttt{Tok} bar). This overhead becomes more pronounced with larger batches or longer contexts, underscoring the inefficiency of full recomputation. Storing intermediate activations instead of recomputing them could mitigate this cost by bypassing redundant FFN processing.

\subsection{Potentials of Activation Cache}
We apply an \underline{\textit{Activation cache}} as a novel alternative to the conventional KV cache, aiming to enhance KV recomputation efficiency via \underline{\textit{activation checkpointing}}. By storing only the decoder input activations, our approach allows faster KV regeneration at each layer, skipping redundant operations after QKV generation without sacrificing accuracy. 
As a result, Activation cache provides the potential of a memory-efficient and computationally balanced framework for large-scale LLM serving with host memory offloaded system.

\begin{figure}[t]
    \centering
    \includegraphics[width=0.95\linewidth]{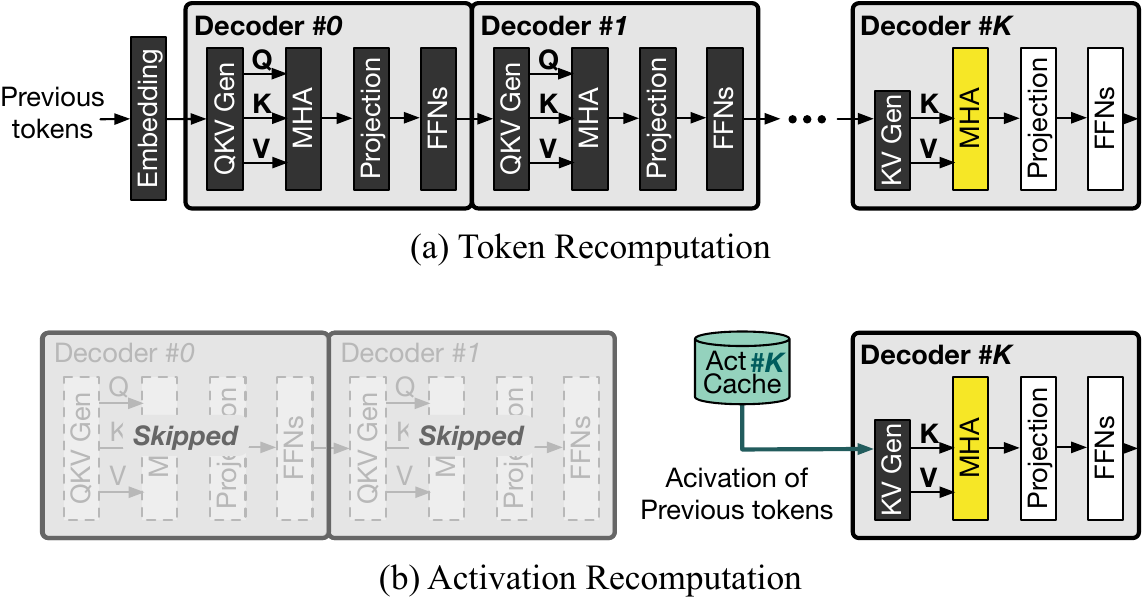}
    \caption{Computation difference for retrieving context of previous tokens for multi-head attention, in (a) Token recomputation and (b) Activation recomputation.}
     \vspace{-2ex}
    \label{fig:cache-difference}
\end{figure}

\begin{figure}[t]
    \centering
    \includegraphics[width=0.9\linewidth]{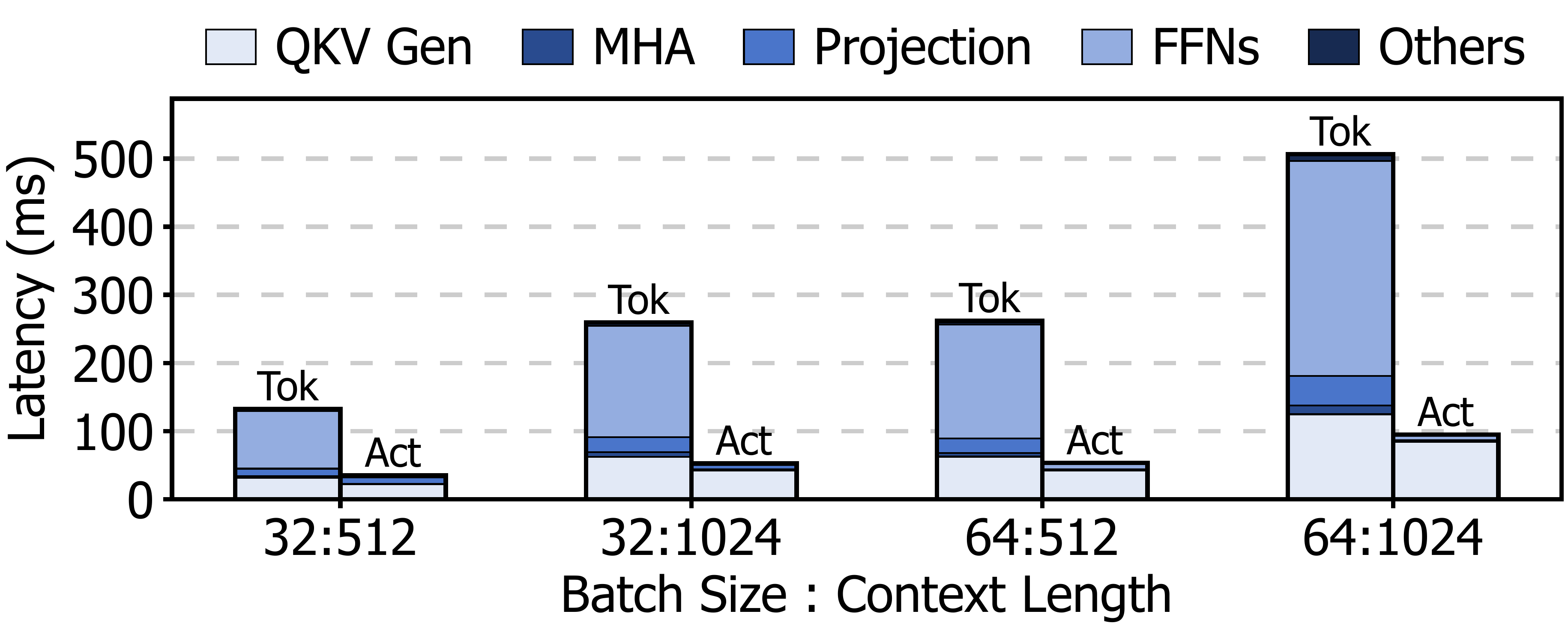}
    \caption{Breakdown of a single layer execution time of OPT-30B, with token recomputation from token IDs (Tok) and activation recomputation from activation checkpoints (Act).}
    \label{fig:remat-breakdown}
\end{figure}

\niparagraph{Efficiency over KV cache.}
Unlike KV caching, which stores both key and value tensors, our Activation cache retains only the input activations ($\mathbf{A_c}$). As shown in Equation~\ref{eq:act-cache-recompute}, these activations are converted into $\mathbf{K_c}$ and $\mathbf{V_c}$ through a single linear transformation.
\begin{equation}
[\mathbf{K_c} \quad \mathbf{V_c}] = \mathbf{A_c} \times [\mathbf{W_{K}} \quad \mathbf{W_{V}}]
\label{eq:act-cache-recompute}
\end{equation}
This reduces memory and data transfer overhead by 50\%. Furthermore, activation checkpointing often leverages idle GPU cycles for recomputation, balancing communication and compute demands more effectively than KV caching.

\niparagraph{Efficiency over token recomputation.}
Compared to token recomputation that accompanies computations of all preceding layers to rebuild KV tensors for layer $k$ (Figure~\ref{fig:cache-difference}(a), activation checkpointing regenerates $\mathbf{K_c}$ and $\mathbf{V_c}$ only from the stored activations of layer $k$ (Figure~\ref{fig:cache-difference}(b)). 
In Figure~\ref{fig:remat-breakdown}, the single-layer execution latency of OPT-30B~\cite{opt} is reduced by an average of 78\% when using activation checkpointing for KV cache recomputation (\texttt{Act}) instead of token recomputation (\texttt{Tok}), highlighting a substantial computational advantage.

\section{\capture System Design}
\label{sec:capture-design}

\begin{figure*}[t!]
    \centering
    \includegraphics[width=1.0\textwidth]{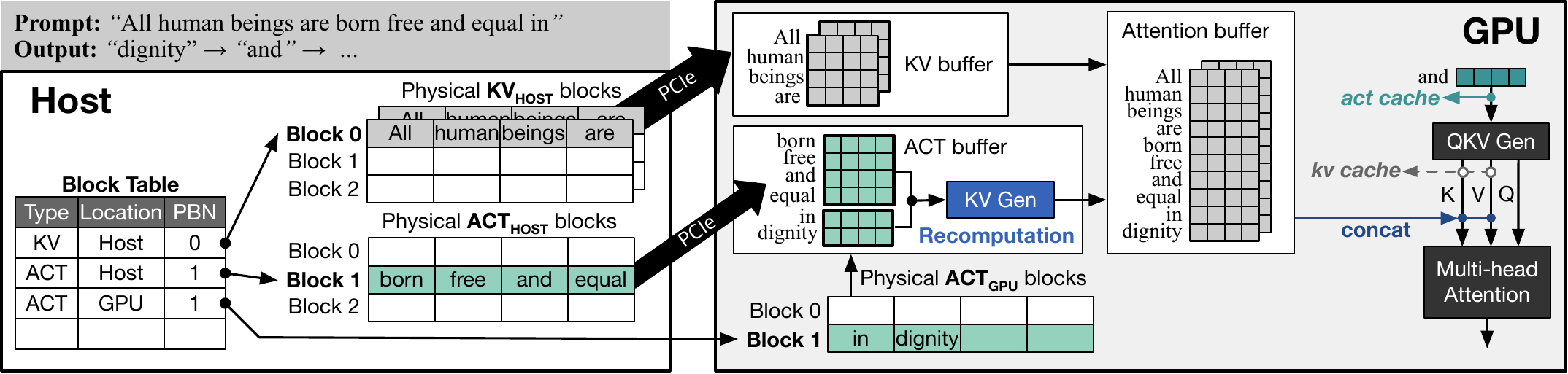}
    \vspace{-0.1in}
    \caption{Overview of the proposed system with KV-Activation hybrid caching mechanism.}
    \vspace{-0.1in}
    \label{fig:hybrid-cache}
\end{figure*}

Replacing the entire KV cache with an activation cache can reduce memory usage, but at the cost of increased recomputation. Building on these insights, we present \capture, a host-memory offloading framework designed to optimize both memory efficiency and computational performance. \capture incorporates \textit{KV-Activation hybrid caching} to balance storage and recomputation, an \textit{asynchronous inference engine} to overlap data transfers with GPU execution, and a \textit{cache management policy} to dynamically allocate resources and minimize bottlenecks.

\subsection{KV-Activation Hybrid Caching}
\niparagraph{Hybrid cache block structure.}
We propose a \emph{hybrid caching} mechanism that divides tokens into two distinct block types: \emph{KV blocks}, which store conventional key-value (KV) tensors, and \emph{ACT blocks}, which store input activations (i.e., activation checkpoints). Each block represents a group of 16 tokens, analogous to the concept of a ``page” in PagedAttention\cite{pagedattention}. As depicted on the left side of Figure~\ref{fig:hybrid-cache}, these KV and ACT blocks are jointly managed by a unified block table that spans host and GPU memory. As the number of tokens per block is fixed, each KV block requires twice the memory space of an ACT block. This design offers a balanced trade-off between memory usage and computational overhead: while KV blocks facilitate rapid attention computations, ACT blocks enable on-demand KV regeneration directly on the GPU, thereby reducing both storage requirements and data transfer costs.

\niparagraph{Logical-Physical block mapping.}
To accommodate unpredictable LLM output lengths, \capture adopts PagedAttention's page-based strategy that enables non-contiguous physical memory allocation while maintaining logical contiguity, thereby reducing the internal fragmentation caused by traditional maximum-size memory reservation approaches.
During the prefill phase, tokens from each request are stored either in KV or ACT blocks, tracked by a per-request block table (Figure~\ref{fig:hybrid-cache}, left) that records block type, physical location, and Physical Block Number (PBN). During inference, \capture consults this table to either directly load KV blocks or regenerate them from ACT blocks. By dynamically tuning the ratio of KV to ACT blocks, the system can effectively overlap PCIe transfers and computations.

\subsection{Asynchronous Inference Engine}
\niparagraph{Three-step decode pipeline.}
We design an \emph{asynchronous inference engine} that overlaps host-to-GPU data transfers with computations. As shown on the right side of Figure~\ref{fig:hybrid-cache}, each iteration of the decode phase consists of three steps: loading activation or KV blocks into GPU buffers, recomputing KVs from activation checkpoints (``KV Gen"), and executing the forward pass for the newly generated token.

\niparagraph{Memory allocation strategy.}
A key design choice is that while host memory can store both Activation cache and KV cache, GPU memory is exclusively allocated for Activation cache blocks. This allocation provides sufficient time for KV recomputation from activation checkpoints during weight prefetching, making optimal use of limited GPU memory during token generation.

\niparagraph{Double-buffered execution flow.}
Our engine employs double-buffering to overlap data transfers with GPU computations. During each decode iteration, three operations occur in parallel: (1) activation blocks are loaded and converted to KV tensors on the GPU, (2) the next layer's weights and existing KV blocks are prefetched from host memory, and (3) the regenerated and prefetched KV tensors are merged for attention computation. This pipeline repeats for each token throughout both the prefill and decode phases, ensuring continuous computation-communication overlap.

\niparagraph{Mini-batch processing and buffer management.}
Our scheduler splits large batches into mini-batches of up to 8,192 tokens to prevent memory bloating from intermediate tensors during attention computation. Without this constraint, large batch sizes would overflow GPU memory with temporary matrices. We manage these mini-batches through three specialized GPU buffer pairs: the \emph{ACT buffer} for activation blocks, the \emph{KV buffer} for key-value blocks from host memory, and the \emph{Attention buffer} for merged KV cache. Each buffer type is duplicated for double-buffering—while one buffer is used to process the current mini-batch on the GPU, the other buffer is used to receive the next mini-batch from the host. With buffer capacity matching the maximum mini-batch size (8,192 tokens), this design prevents memory bloating while enabling continuous processing across mini-batches.

\begin{figure}[t]
    \centering
    \includegraphics[width=0.9\linewidth]{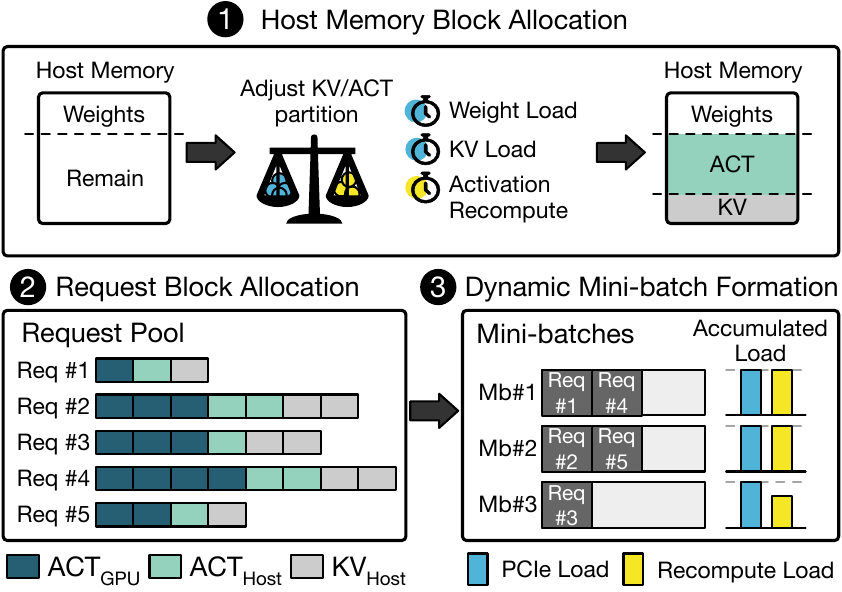}
    \caption{Overview of hybrid cache management policy.}
    \vspace{-2ex}
    \label{fig:scheduling-overview}
\end{figure}

\subsection{Cache Management Policy}
Efficient inference requires a seamless overlap between PCIe data transfers and GPU computation. Over-allocating KV cache increases PCIe traffic, resulting in GPU idle time. Conversely, overly relying on activation caching leads to frequent KV regeneration, leaving the PCIe bandwidth underutilized. Therefore, the key challenge is dynamically balancing the ACT-to-KV cache ratio to achieve optimal resource utilization.

To address this challenge, we propose a cache allocation and scheduling policy for KV-Activation hybrid cache. As illustrated in Figure~\ref{fig:scheduling-overview}, our policy consists of three key steps: \niparagraph{\circled{1} Host memory block allocation} determines cache allocation between KV and ACT blocks in host memory. \niparagraph{\circled{2} Request block allocation} assigns blocks to each request based on the host memory allocation ratio. \niparagraph{\circled{3} Dynamic mini-batch formation} organizes mini-batches to balance PCIe communication and recomputation.

\begin{table}[t]
    \centering
    \footnotesize
    \begin{tabular}{ll}
        \toprule
        \multicolumn{2}{c}{\textbf{Terms \& Functions for Time}}\\
        \toprule
        $T_{load\_w}$ & Latency for one decoder block weight load \\
        $T_{kv\_gen}$ & Function for KV generation (recompute) latency \\
        $T_{load\_kv}$ & Function for KV cache load latency \\
        \toprule
        \multicolumn{2}{c}{\textbf{Terms for \# of Blocks}}\\
        \toprule
        \#$KV_{Host}$ & Number of KV blocks in host memory \\
        \#$ACT_{Host}$ & Number of ACT blocks in host memory \\
        \#$ACT_{GPU}$ & Number of ACT blocks in GPU memory \\
        \toprule
    \end{tabular}
    \caption{Symbols for allocation and scheduling decision.}
    \label{tab:scheduling-reference}
    \vspace{-4ex}
\end{table}

\niparagraph{Problem definition.}
We formalize the inference pipeline latency for a single Transformer decoder layer using the notation defined in Table~\ref{tab:scheduling-reference}. The latency consists of two main components: (i) $T_{PCIe}$, the time required to load weights and KV blocks from host memory, and (ii) $T_{Computation}$, the time needed to regenerate KV cache from ACT blocks stored in GPU and host memory.
These latencies are defined as:
\begin{gather}
T_{PCIe} = T_{load\_w} + T_{load\_kv}(\text{\#}KV_{Host}) \label{eq:t-pcie} \\[1ex]
T_{Computation} = T_{kv\_gen}(\text{\#}ACT_{Host} + \text{\#}ACT_{GPU}) \label{eq:t-compute}
\end{gather}

Our objective is to balance these two latencies, avoiding prolonged idle time for either the GPU or the PCIe bus. This can be expressed as minimizing:
\begin{gather}
\text{Minimize } | T_{PCIe} - T_{Computation} |
\label{eq:balance-problem}
\end{gather}

\subsubsection{Host Memory Block Allocation}
Host memory allocation between KV and ACT blocks directly determines the balance between PCIe traffic and computational workload, as shown in equations (\ref{eq:t-pcie}) and (\ref{eq:t-compute}). Our strategy employs a two-step approach (Figure~\ref{fig:scheduling-overview} \circled{1}, Algorithm~\ref{alg:allocation}): First, we establish an initial allocation to eliminate resource idle time by leveraging the natural overlap between activation recomputation and weight loading. Second, we distribute the remaining host memory capacity to optimize the balance between PCIe and computation latencies.

\niparagraph{Initial allocation.}
Initially, the number of host blocks required to eliminate idle time is calculated by comparing the recomputation time of ACT blocks allocated to the GPU with the time taken to load weights (lines 11-18 of Algorithm~\ref{alg:allocation}).
If weight loading time exceeds recomputation time, additional host ACT blocks ($ACT_{init}$) are allocated to prevent GPU idle time. Conversely, if computation time is longer, host KV blocks ($KV_{init}$) are prefetched to prevent PCIe idle time.

\niparagraph{Remaining allocation.}
After establishing the baseline allocation, we distribute the remaining host memory capacity between additional activation and KV blocks to achieve optimal latency balance. The remaining capacity is calculated by subtracting the memory required for weight parameters and baseline blocks from the total host memory. We then determine the number of additional ACT and KV blocks such that $T_{kv\_gen}(\text{\#}ACT) = T_{load\_kv}(\text{\#}KV)$, ensuring a balance between the two latencies. Since both latencies scale linearly with block count ($T_{kv\_gen}\propto\text{\#}ACT$, $T_{load\_kv}\propto\text{\#}KV$) and $S_{ACT} = \frac{1}{2}S_{KV}$, this optimization reduces to solving a straightforward linear system.

\begin{algorithm}[t]
\caption{Hybrid Cache Allocation Policy}
\label{alg:allocation}
\begin{algorithmic}[1]
    \small
    \Statex \hskip-1.5em\textbf{Parameter:} $T_{load\_w}$: Weight load latency for one decoder block
    \Statex \hskip3em $\text{\#}ACT_{GPU}$: Number of ACT blocks in GPU memory
    \Statex \hskip3em $M_{Host}$: Total capacity of host memory
    \Statex \hskip3em $S_{weight}$: Total size of weight parameter
    \Statex \hskip3em $S_{KV}$: Size of a KV block
    \Statex \hskip3em $S_{ACT}$: Size of an ACT block $(=\frac{1}{2}S_{KV})$
    \Statex \hskip-1.5em\textbf{Output:} $\text{\#}ACT\_{Host}$, $\text{\#}KV\_{Host}$
    \Statex
    \State{\textit{\# Step 1: Initial allocation}}
    \State $ACT_{init}, KV_{init} \leftarrow\ $     \texttt{init\_cache\_alloc}()
    \State
    \State{\textit{\# Step 2: Allocate remaining blocks after initial alloc}}
    \State $ACT_{remain}, KV_{remain} $ 
    \State \hskip1.5em $ \leftarrow\ $ \texttt{alloc\_remain}($ACT_{init}, KV_{init}$)
    \State
    \State{\textit{\# Step 3: Return } $\text{\#}ACT_{Host}$, $\text{\#}KV_{Host}$}
    \State \textbf{return} ($ACT_{init} + ACT_{remain}), (KV_{init} + KV_{remain})$
    \State
    \State \textbf{def} \texttt{init\_cache\_alloc}():
    \State \hskip1.5em $ACT_{init}, KV_{init} \leftarrow\ 0, 0$
    \State \hskip1.5em $T_{budget} = T_{load\_w} - T_{kv\_gen}(\text{\#}ACT_{GPU}) $ \label{line:t-budget}
    \State \hskip1.5em \textbf{if} $T_{budget} \geq 0$ \textbf{then}
    \State \hskip3em $ACT_{init} \leftarrow\ $ find $\text{\#}ACT$ s.t. $T_{kv\_gen} = T_{budget}$ \label{line:init-act}
    \State \hskip1.5em \textbf{else}
    \State \hskip3em $KV_{init} \leftarrow\ $ find $\text{\#}KV$ s.t. $T_{load\_kv} = -T_{budget}$ \label{line:init-kv}
    \State \hskip1.5em \textbf{return} $ACT_{init}, KV_{init}$
    \State
    \State \textbf{def} \texttt{alloc\_remain}($ACT_{init}, KV_{init}$):
    \State \hskip1.5em $M_{occupied} = S_{ACT} \times ACT_{init} + S_{KV} \times KV_{init}$  \label{alg:calc-occupied}
    \State \hskip1.5em $M_{remaining} = M_{Host} - S_{weight} - M_{occupied}$ \label{line:m-remain}
    \State \hskip1.5em $ACT_{remain}, KV_{remain} \leftarrow\ $ find $\text{\#}ACT$, $\text{\#}KV$ s.t.
    \State \hskip3em $M_{remaining} = S_{ACT} \times \text{\#}ACT + S_{KV} \times \text{\#}KV$ \label{alg:calc-remaining}
    \State \hskip3em $T_{kv\_gen}(\text{\#}ACT) = T_{load\_kv}(\text{\#}KV)$ \label{alg:linear-system}
    \State \hskip1.5em \textbf{return} $ACT_{remain}, KV_{remain}$
\end{algorithmic}
\end{algorithm}


\subsubsection{Request Block Allocation}
Based on the two-step allocation policy, the number of ACT and KV physical blocks to allocate in host memory is set to $ACT_{init} + ACT_{remain}$ and $KV_{init} + KV_{remain}$, respectively. Since this ratio balances $T_{PCIe}$ and $T_{Computation}$, \capture ensures that each request maintains the same ratio when storing its KV cache, as shown in Figure~\ref{fig:scheduling-overview} \circled{2}:
\begin{equation}
\text{\#}ACT_{req} : \text{\#}KV_{req} = \text{\#}ACT_{Host} : \text{\#}KV_{Host}
\label{eq:block-ratio}
\end{equation}
$\text{\#}ACT_{req}$ and $\text{\#}KV_{req}$ represent the number of each block type that a request uses. After the prefill phase, context blocks are stored through either activation checkpointing or KV caching according to this ratio. During the decode phase, new blocks are allocated maintaining this ratio.
For instance, if the ratio of $\text{\#}ACT_{Host}$ and $\text{\#}KV_{Host}$ is 3:1 when a request has five ACT blocks and two KV blocks, the next block allocated would be an ACT block.

\subsubsection{Dynamic Mini-batch Formation}
\label{subsec:bin-packing}
While Algorithm~\ref{alg:allocation} ensures a balance between KV and activation within a single request, it is also important to consider this balance at the mini-batch level for better performance. Since a mini-batch handles both computation and communication, balancing KV and activation dynamically is essential. This approach not only maximizes PCIe and computational resource utilization, but also improves the efficiency of double buffering.

\niparagraph{Ensuring pipeline balance.}
To achieve the best performance, mini-batch should always keep the balance between $T_{kv\_gen}$ and $T_{load\_kv}$.
We use $balance$ as a metric to measure how balanced two pipelines are in a mini-batch.
Basically, $balance$ is a ratio of $T_{kv\_gen}$ and $T_{load\_kv}$ in the current mini-batch, as defined in Equation~\ref{eq:balance}.
\begin{gather}
    balance = T_{kv\_gen}(\text{\#}ACT_{mb})/T_{load\_kv}(\text{\#}KV_{mb})
    \label{eq:balance}
\end{gather}
Note that $\text{\#}KV_{mb}$ and $\text{\#}ACT_{mb}$ are the number of KV blocks and ACT blocks for the current mini-batch.
Since an ideal value of $balance$ is 1, the purpose of the request scheduling algorithm is to make $balance$ converge to 1 in each mini-batch.
To transform this into the minimization problem, we define a cost function $F_b$ as Equation~\ref{eq:balance-reverse}.
\begin{equation}
    F_b(\text{\#}ACT_{mb}, \text{\#}KV_{mb}) = max(balance, 1/balance)
    \label{eq:balance-reverse}
\end{equation}
The goal of our mini-batch formation algorithm is minimizing $F_b$ for each mini-batch.

\niparagraph{Solving bin packing problem for mini-batches.}
Our dynamic mini-batch formation algorithm uses a greedy approach to minimize both the number of mini-batches and the imbalance metric $balance$. The algorithm first defines $\text{\#}ACT_{max}$ and $\text{\#}KV_{max}$ based on the available GPU buffer size, which serve as capacity constraints for each mini-batch (bin). For each request, the algorithm checks two conditions: (1) whether the request fits within the remaining capacity constraints ($\text{\#}KV$ and $\text{\#}ACT$), and (2) whether adding the request reduces the imbalance compared to the current mini-batch state. If both conditions are satisfied, the request is added to the current mini-batch. When no more requests can be accommodated, a new mini-batch is created. This process continues until all requests are allocated, ensuring efficient resource utilization while maintaining balance.

\section{Evaluation}

\begin{figure*}[t]
    \centering
    \includegraphics[width=\linewidth]{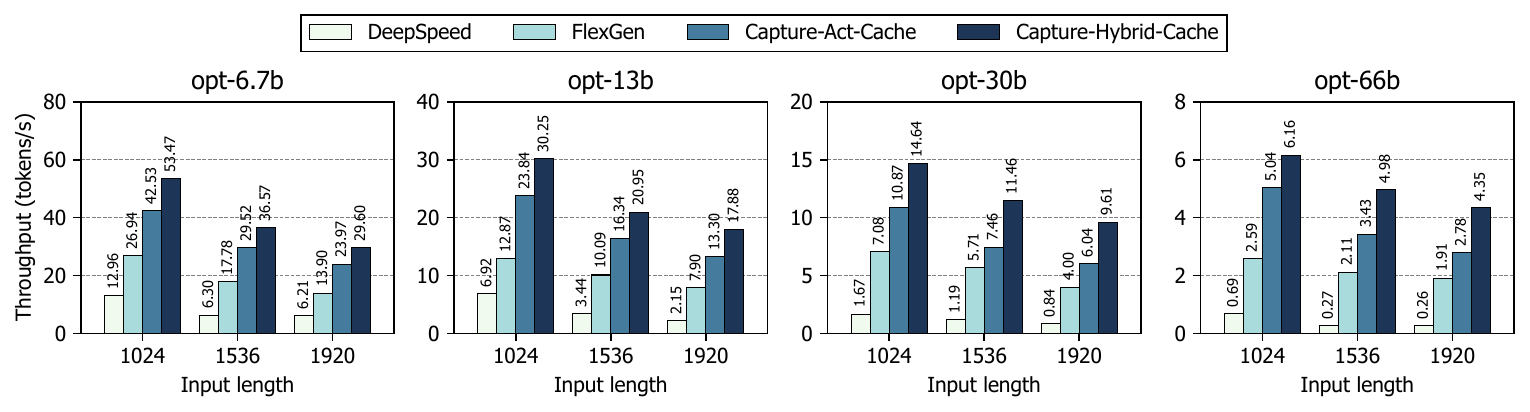}
    \vspace{-3ex}
    \caption{Throughput of \capture with various OPT model sizes.}
    \vspace{-2ex}
    \label{fig:throughput-improvement}
\end{figure*}

\begin{figure}[t]
    \centering
    \includegraphics[width=1.0\linewidth]{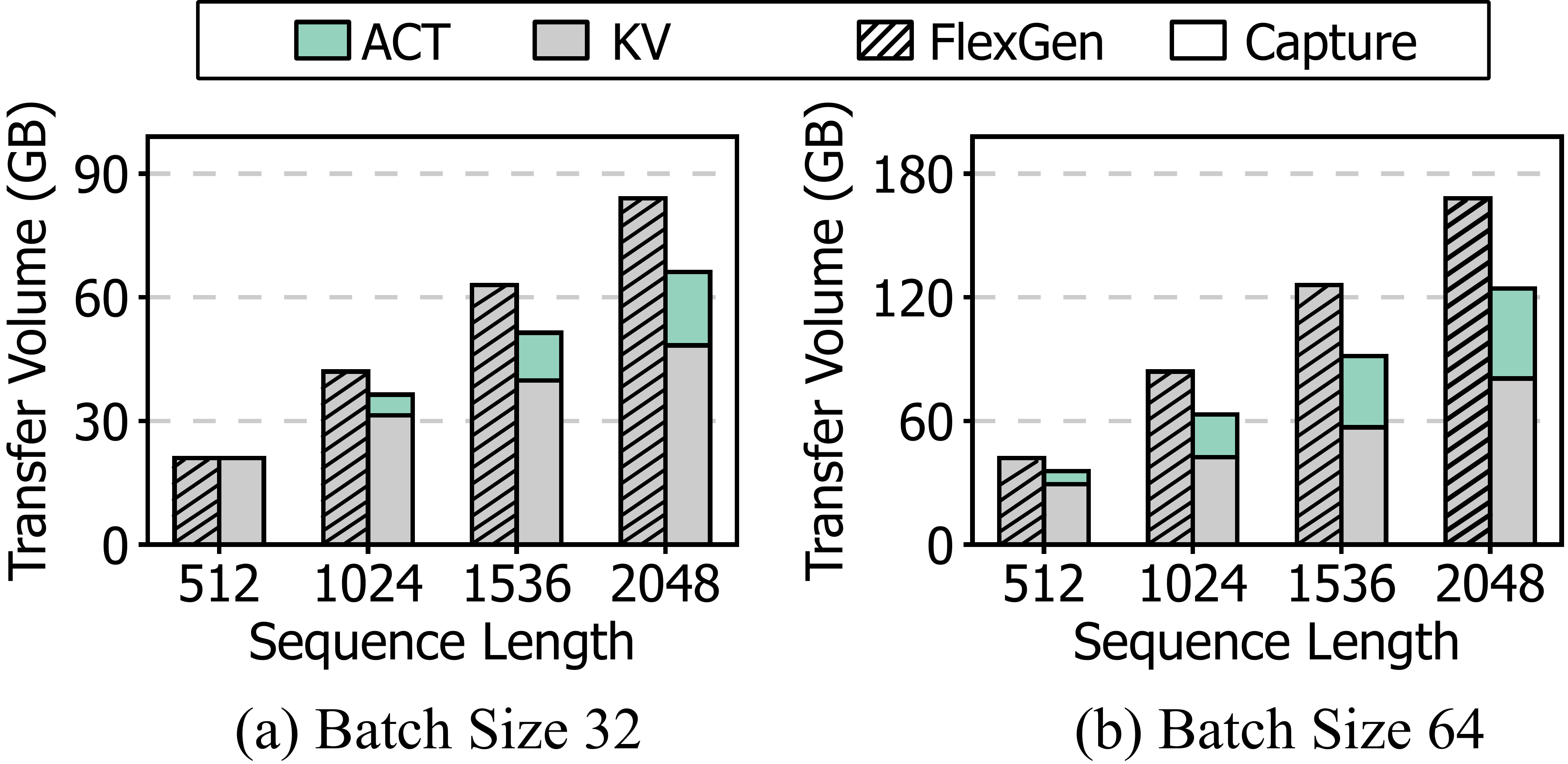}
    \caption{Breakdown of PCIe transfer volume for KV and Activation using OPT-30B. Shaded bars represent FlexGen, while solid bars represent \capture.}
    \vspace{-3ex}
    \label{fig:transfer-volume}
\end{figure}

\subsection{Methodology}

\niparagraph{Implementation.}
\capture is a single-GPU LLM inference system that extends vLLM~\cite{pagedattention}, incorporating host memory offloading capabilities for both weights and KV caches. 
\emph{KV-Activation hybrid caching} is implemented by modifying 12K lines of code using Python and C++/CUDA with PyTorch.
To ensure seamless integration of the hybrid cache, \capture expands PagedAttention kernel of vLLM to enable self-attention functionality across diverse KV cache types.

\niparagraph{Models.}
We evaluate \capture with four OPT~\cite{opt} model sizes: 6.7B, 13B, 30B, and 66B parameters (all in float16). While larger models (13B-66B) require host memory offloading, OPT-6.7B fits entirely in GPU memory. We include it to show how activation caching reduces PCIe traffic compared to KV caching, even for models not requiring offloading.

\niparagraph{Baselines.}
We compare \capture with two modern host memory offloading frameworks: DeepSpeed-Inference~\cite{deepspeed-inference} and FlexGen~\cite{flexgen}. We exclude sparsification-based frameworks like InfiniGen~\cite{infinigen}  that have potential negative impacts on language modeling accuracy, though these methods could complement our approach by reducing data transfer.
For FlexGen, we use its optimal configuration, which keeps most weights on the GPU while offloading KV cache and remaining weights to host memory. DeepSpeed-Inference, lacking zig-zag scheduling, uses the batch size that avoids OOM during the prefill phase, resulting in smaller batches compared to FlexGen. We also evaluate \capture-Act-Cache, a variant of \capture using only activation caching.

\niparagraph{Environment Setup.}
We perform evaluation in a single GPU system featuring an NVIDIA RTX 4090 GPU with 24GB of GDDR6X memory and PCIe 4.0 x16 interface.
The host system is powered by a dual-socket 16-core Intel Xeon Gold 6326 processor, with 882GB of DDR4 memory.

\niparagraph{Evaluation metrics.}
We evaluate \capture based on token generation throughput, defined as the number of tokens generated per unit time, a critical metric for batched LLM inference. Additionally, we analyze KV transfer volume between the host and GPU and measure GPU temporal utilization, defined as the percentage of cycles with active computation, using NVIDIA Nsight Systems.

\subsection{Throughput Improvement}

Figure~\ref{fig:throughput-improvement} compares the throughput of \capture (\capture-Hybrid-Cache) with FlexGen, DeepSpeed-Inference, and Activation-cache only system (\capture-Act-Cache) across varying input prompt lengths. Throughput is calculated as the total number of tokens divided by the end-to-end latency (prefill + decode), with a batch size of 128 and 128 output tokens per request.

\niparagraph{Comparison with KV cache-only systems.}  
\capture-Hybrid-Cache improves throughput by 2.19× over FlexGen and 1.61× over \capture-Act-Cache by efficiently utilizing GPU cycles for KV recomputation, reducing host-to-GPU KV cache traffic and alleviating data transfer bottlenecks. These benefits scale with model size; for example, throughput improves by 2.05× for 6.7B models and 2.33× for 66B models, where longer weight transfer times enable greater overlap with activation recomputation.

In contrast, DeepSpeed-Inference achieves only 29\% of FlexGen’s throughput and 13\% of \capture-Hybrid-Cache’s due to its lack of mini-batch scheduling. Without the ability to divide large batches into smaller mini-batches, it is forced to use smaller batch sizes in memory-constrained scenarios, significantly limiting throughput. Prior studies~\cite{flexgen} have reported similar scalability challenges in DeepSpeed-Inference.

\niparagraph{Comparison with Activation cache-only systems.}  
\capture-Hybrid-Cache improves throughput by an average of 1.35× over \capture-Act-Cache by alleviating both data transfer and computational bottlenecks. While an activation cache-only system eliminates host-to-GPU traffic, it incurs excessive KV regeneration, straining GPU resources. Hybrid caching mitigates this by dynamically balancing KV and activation blocks, optimizing both memory efficiency and computational overhead. This advantage is more pronounced in larger models, where regeneration costs scale with model size. Specifically, hybrid caching achieves a 1.23× speedup for the 6.7B model and 1.4× for the 66B model, demonstrating its effectiveness in sustaining high throughput as model complexity increases.

\subsection{Host-GPU Traffic Analysis}
Figure~\ref{fig:transfer-volume} presents the host-to-GPU data transfer breakdown during decoding across different input lengths for batch sizes of (a) 32 and (b) 64. The x-axis represents input sequence length, while the y-axis indicates the transfer volume. For each sequence length, the left bar (comb-patterned) represents FlexGen, while the right bar corresponds to \capture.

Replacing KV blocks with smaller ACT blocks, \capture reduces traffic by up to 1.27× for batch size 32 and 1.38× for batch size 64 compared to FlexGen. This advantage scales with batch size, as larger batches require transferring proportionally more KV cache over PCIe. By substituting some KV cache entries with Activation cache, \capture mitigates PCIe bottlenecks in high-throughput settings.

\begin{figure}[t]
    \centering
    \includegraphics[width=1.0\linewidth]{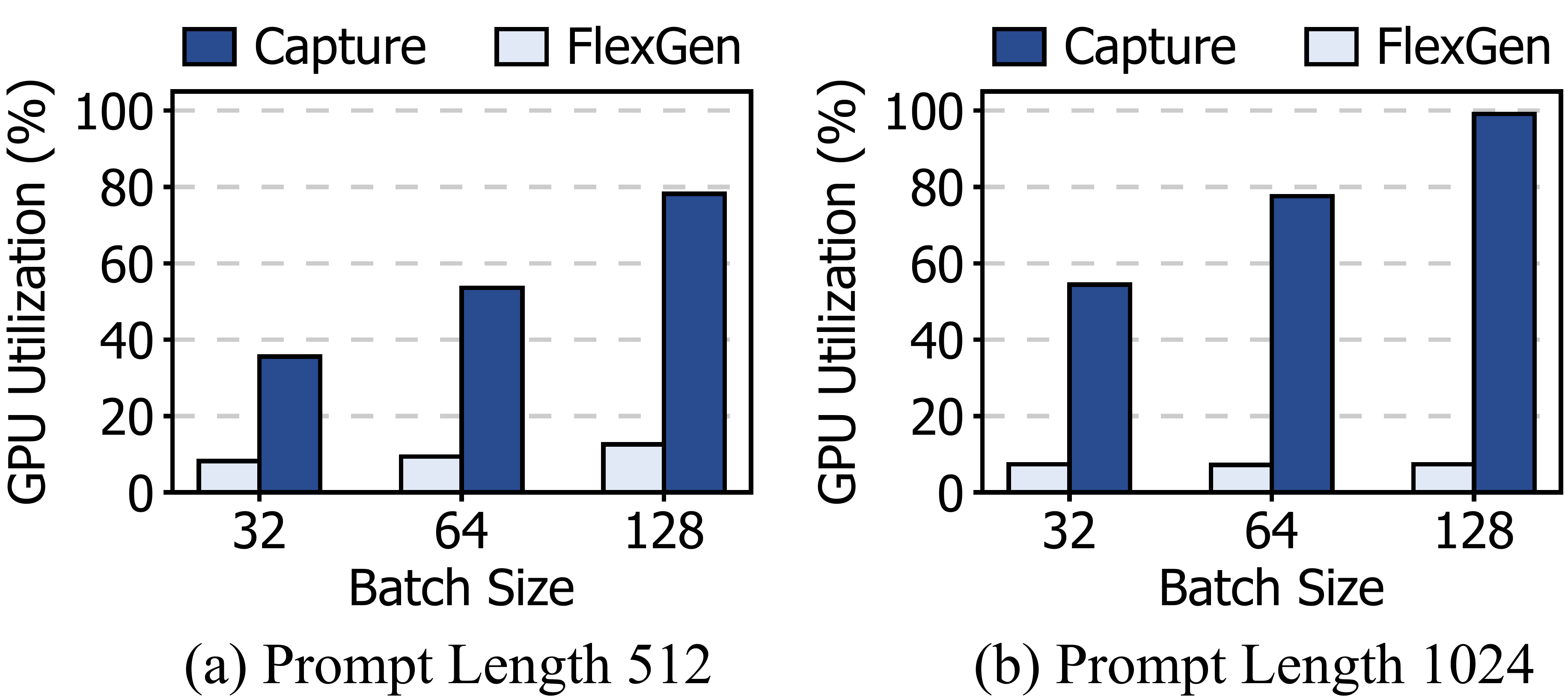}
    \caption{GPU utilization of OPT-30B with varying batch sizes and input prompt lengths, comparing FlexGen and \capture.}
    \label{fig:gpu-utilization}
\end{figure}

\subsection{GPU Utilization}
Figure~\ref{fig:gpu-utilization} compares GPU utilization between FlexGen and \capture across different batch sizes and input lengths. On average, \capture achieves 7.39$\times$ higher utilization than FlexGen, with the largest gain at a batch size of 128, where utilization improves by 13.39$\times$. This disparity occurs because FlexGen heavily depends on KV cache transfers, leading to only a slight increase in GPU utilization from 8.2\% to 12.6\% as batch size grows from 32 to 128.

In contrast, \capture increases GPU utilization from 35.6\% to 78.2\% over the same range. This improvement is driven by its ability to generate activation recomputation tasks dynamically, leveraging available GPU resources while minimizing idle time. While higher utilization does not guarantee proportional throughput gains, \capture reduces KV cache traffic and enhances resource efficiency, making it more effective than FlexGen in maximizing GPU performance.

\subsection{Effect of Cache Management Policy}
\begin{table}[t]
    \centering
    \small
    \setlength{\tabcolsep}{6pt} 
    \begin{tabular}{r|c c c}
        \toprule
        \textbf{\#Params} & \textbf{Act-Only} & \textbf{+ Hybrid} & \textbf{+ Alloc \& Sched} \\
        \midrule
        6.7B  & 25.06 & 31.49 \scriptsize{(1.26×)} & \textbf{31.53 \scriptsize{(1.26×)}} \\
        13B   & 13.90 & 18.61 \scriptsize{(1.34×)} & \textbf{19.06 \scriptsize{(1.37×)}} \\
        30B   & 6.28  & 8.68  \scriptsize{(1.38×)} & \textbf{10.10 \scriptsize{(1.61×)}} \\
        66B   & 2.96  & 4.12  \scriptsize{(1.39×)} & \textbf{4.63  \scriptsize{(1.56×)}} \\
        \bottomrule
    \end{tabular}
    \caption{Ablation study showing throughput (tokens/s) for activation cache, hybrid cache, and hybrid cache with allocation and scheduling, with speedups over activation cache in parentheses.}
    \vspace{-3ex}
    \label{tab:ablation}
\end{table}

Table~\ref{tab:ablation} presents an ablation study on the impact of hybrid caching and cache management policies on throughput, using a prompt length of 1920 and a batch size of 128. Without dynamic scheduling and allocation, hybrid caching follows a static policy where Activation cache is filled first, followed by KV cache. This enforces a fixed 1:1 host memory split between Activation and KV cache, allocating equal host memory to both. While this approach improves throughput by an average of 1.34×—with gains of 1.26× for the 6.7B model and 1.39× for the 66B model—its effectiveness is limited. Larger models see greater benefits due to longer weight loading times, which allow more activation recomputation. However, static allocation does not account for pipeline imbalances, causing recomputation to exceed PCIe latency and restricting further speedups.

Introducing cache management policies, including dynamic host memory allocation and bin-packing, further improves throughput—achieving speedups of 1.61× for the 30B model and 1.56× for the 66B model compared to \capture-Act-Cache. For smaller models, these improvements are marginal, as their default 1:1 memory split aligns with the optimal ratio for overlap. In contrast, larger models benefit significantly, with optimal KV-to-ACT memory ratios of 2:1 (30B) and 1.78:1 (66B). By dynamically adjusting memory allocation and improving overlap efficiency, these policies mitigate bottlenecks and enhance overall performance.
\section{Conclusion}


This paper addresses communication-computation imbalance in host memory offloading for batched LLM inference with activation checkpointing and KV-Activation hybrid caching. This improves throughput and efficiency, enabling scalable, cost-effective inference. 
The source code of \capture framework is available at \url{https://github.com/casys-kaist/Capture}.
\section{Acknowledgements}
This work was supported by National Research Foundation of Korea (NRF; RS-2024-00347114), and Institute of Information \& communications Technology Planning \& Evaluation (IITP) funded by the Ministry of Science and ICT, Korea (RS-2021-II211817, IITP-2024-RS-2023-00256472).

\bibliographystyle{IEEEtran}
\bibliography{reference}

\end{document}